\documentclass[12pt,preprint]{aastex}
\usepackage{epsfig}
\usepackage{lscape}
\def\spose#1{\hbox to 0pt{#1\hss}}
\def\simlt{\mathrel{\spose{\lower 3pt\hbox{$\mathchar"218$}}
     \raise 2.0pt\hbox{$\mathchar"13C$}}}
\def\simgt{\mathrel{\spose{\lower 3pt\hbox{$\mathchar"218$}}
     \raise 2.0pt\hbox{$\mathchar"13E$}}}

\begin{document}
\title{On the typical timescale for the chemical enrichment from SNeIa 
in galaxies}
\author{Francesca Matteucci and Simone Recchi}
\affil{Dipartimento di Astronomia, Universit\`a di Trieste, via G.B. 
Tiepolo 11, 34131 Trieste, Italy}

\begin{abstract}
We calculate the type Ia supernova rate for different star formation
histories in galaxies by adopting the most popular and recent
progenitor models. We show that the timescale for the maximum in the
type Ia supernova rate, which corresponds also to time of the maximum
enrichment, is not unique but is a strong function of the adopted
stellar lifetimes, initial mass function and star formation rate. This
timescale varies from $\sim\;40-50$ Myr for an instantaneous starburst
to $\sim$ 0.3 Gyr for a typical elliptical galaxy to $\sim\;4.0-5.0$
Gyr for a disk of a spiral Galaxy like the Milky Way. We also show
that the typical timescale of 1 Gyr, often quoted as the typical
timescale for the type Ia supernovae, is just the time at which, in
the solar neighbourhood, the Fe production from supernovae Ia starts
to become important and not the time at which SNe Ia start to explode.
As a cosequence of this, a change in slope in the [O/Fe] ratio is
expected in correspondance of this timescale.  We conclude that the
suggested lack of supernovae Ia at low metallicities produces results
at variance with the observed [O/Fe] vs. [Fe/H] relation in the solar
region. We also compute the supernova Ia rates for different galaxies
as a function of redshift and predict an extended maximum between
redshift $z \sim 3.6$ and $z \sim 1.6$ for elliptical galaxies, and
two maxima, one at $z \sim 3$ and the other at $z \sim 1$, for spiral
galaxies, under the assumption that galaxies start forming stars at
$z_f \sim 5$ and $\Omega_M = 0.3$, $\Omega_{\Lambda} = 0.7$.
\end{abstract}


\section{INTRODUCTION}

The supernovae of type Ia are crucial in understanding a number of
astrophysical problems of primary importance, such as the supernova
(SN) progenitors, the determinations of cosmological constants, the
chemical enrichment of galaxies and the thermal history of the
interstellar (ISM) and intracluster (ICM) medium. In the past year
there has been a great deal of work concerning the different roles of
different types of SNe on the evolution of galaxies.

Tinsley (1980) first discussed the possibility of iron being mostly
produced in SNe of type I (the distinction in type Ia and Ib was not
common at that time) and in the following years type Ia supernovae
started to be considered as important iron contributors in the
chemical evolution of galaxies (Greggio and Renzini 1983a,b; Matteucci
and Greggio 1986). In particular, the observed [O/Fe] versus [Fe/H]
relation was interpreted as due to the different roles played by SNe
of type II (explosion of single massive stars) and type Ia in the
production of oxygen and iron. Supernovae of type II are, in fact,
responsible for the bulk of oxygen and for part of iron production
whereas type Ia are responsible for a large fraction of iron
production, the exact value depending on the assumed initial mass
function.

At the same time models for the progenitors of SNe of type Ia were
developed (Iben and Tutukov 1984; 1985). Such models suggest that
these SNe may originate from white dwarfs in binary systems, either
systems with two white dwarfs or systems with one white dwarf and a
red (super) giant, as originally proposed by Whelan and Iben
(1973). In parallel, models of explosive nucleosynthesis for such
systems (mostly the C-deflagration/detonation of a C-O white dwarf of
Chandrasekhar mass) were also presented: the most famous of these
models being the W7 model of Nomoto et al. (1984). Calculations of the
type Ia SN rates were performed on the basis of these models (Greggio
and Renzini 1983b; Matteucci and Greggio 1986; Matteucci and
Tornamb\'e 1987; Tornamb\'e and Matteucci 1987; Matteucci and
Tornamb\'e 1988) both for our galaxy and for elliptical galaxies. In
particular Tornamb\'e and Matteucci (1987) and Matteucci and
Tornamb\'e (1988) showed that the predicted type Ia rates for our
Galaxy and elliptical galaxies are very different because of the
differences in their histories of star formation. They showed that the
maximum of the type Ia SN rate in ellipticals occurs much earlier than
in the solar neighbourhood. The same result was shown more recently by
Matteucci (1994) who adopted a more sophisticated model, including
dark matter, for the evolution of elliptical galaxies. This result has
strong implications on the definition of the typical timescale for the
chemical enrichment by type Ia SNe. Yoshii et al. (1996) pointed out
that the typical timescale for the enrichment from SN Ia in the solar
vicinity can be identified with the time at which the [O/Fe] ratio
starts changing its slope. Independently of the assumed model for
progenitors of type Ia SNe, these authors identified that this
timescale for the solar neighbourhood is $t_{Ia} \sim 1.5$ Gyr, in
agreement with previous estimates by Matteucci and Fran\c cois
(1992). By coincidence, the value of [Fe/H] at which the [O/Fe] ratio
changes slope coincides with the transition between halo and disk
([Fe/H]$\sim$ -1.0). Therefore, this timescale can also be taken as an
indicator for the duration of the halo-thick disk phase. However,
$t_{Ia}$ cannot be universal, as it is often claimed in the
literature, but it depends strongly on the assumed star formation
history. As a consequence of that, we expect that the [$\alpha$/Fe]
vs. [Fe/H] relationships to be different in galaxies with different
histories of star formation (Matteucci 1991).

We will compute here the typical timescale for the maximum enrichment
from type Ia SNe, $t_{Ia}$, in galaxies with different histories of
star formation (starburst, elliptical and spiral galaxies), defined as
the time of the maximum in the type Ia SN rate. In particular, in the
framework of different star formation histories, we will explore
different progenitor models for type Ia SNe, such as the modification
proposed by Greggio (1996) to the classical rate computed by Greggio
and Renzini (1983b) and the more recent model proposed by Hachisu et
al. (1996; 1999).

In section 2 we recall the most popular models for the progenitors of
type Ia SNe and define $t_{Ia}$. In section 3 we compute the rates of
SNe of type Ia in galaxies of different morphological type. In section
4 we discuss the effects of different type Ia SN rates on galactic
chemical evolution. Finally, in section 5 some conclusions are drawn.

\section{TYPE IA SNe, THEIR PROGENITORS, THEIR TIMESCALES}

\subsection{Supernovae Ia}

The main observational constraints about SNeIa are the lack of
hydrogen lines, the presence of Si\,{\sc ii} lines, together with some
other intermediate-mass elements (Ca\,{\sc ii}, S\,{\sc ii}, O\,{\sc
i}), with a wide range of velocities ($\sim 10^4-3\times 10^4$ Km
s$^{-1}$), which dominate the photospheric spectra near the time of
the maximum light. In the later nebular phase, the spectra are
dominated by iron features, thus indicating that the inner layers
consist of iron-peak elements ($\sim 0.1-1 M_{\odot}$).

Most of the observed SNIa lie in a narrow range of parameters
(spectra, light curves, absolute magnitude peak) and only $\sim$ 10 \%
of the observational sample are conspicuously peculiar (mainly
subluminous).

In the light of these constraints, it is widely accepted that SNeIa
originates from the thermonuclear disruption of a white dwarf (WD)
which accretes material from a companion star in close binary systems,
but there are still a number of uncertainties about the nature of the
companion (main sequence star, red giant or another WD), the nature of
the white dwarf (C-O WD, He WD or O-Ne-Mg WD), the mass reached by the
accretor at the explosion (Chandrasekhar or sub-Chandrasekhar) and the
explosion mechanism itself (deflagration, delayed-detonation or
detonation). However, there is a general consensus in assuming that a
C-deflagration in a C-O WD of Chandrasekhar mass best represents the
characteristics of the majority of the SNe Ia.

\subsection{Progenitors}

We recall here the most common models for the progenitors of type Ia
SNe proposed insofar:

\begin{itemize}

\item The merging of two C-O WDs, due to gravitational wave radiation,
which reach the Chandrasekhar mass and explode by C-deflagration (Iben
and Tutukov 1984; 1985). This is known as double-degenerate (DD)
scenario.

\item The C-deflagration of a Chandrasekhar mass ($\sim 1.4
M_{\odot}$) C-O WD after accretion from a non-degenerate companion
(Whelan and Iben 1973; Munari and Renzini 1992; Kenyon et
al. 1993). This model is known as the single-degenerate (SD) one. The
main problem with this scenario is the narrow range of permitted
values of the mass accretion rate in order to obtain a stable
accretion, instead of an unstable accretion with a consequent nova
explosion and mass loss. In this case, in fact, the WD never achieves
the Chandrasekhar mass. In particular, Nomoto, Thielemann \& Yokoi
(1984) found that a central carbon-deflagration of a WD results for a
high accretion rate ($\dot M \simgt 4 \cdot 10^{-8}\, M_\odot \,{\rm
yr}^{-1}$) from the secondary to the primary star (the WD). They found
that $\sim 0.6 - 0.7\; M_{\odot}$ of Fe plus traces of elements from C
to Si are produced in the deflagration, well reproducing the observed
spectra.

\item A sub-Chandrasekhar C-O WD exploding by He-detonation induced by
accretion of He-rich material from a He star companion (Limongi and
Tornamb\'e 1991).

\item A recent model by Hachisu et al. (1996; 1999) is based on the
classical scenario of Whelan and Iben (1973) (namely C-deflagration in
a WD reaching the Chandrasekhar mass after accreting material from a
star which fills its Roche lobe), but they find an important
metallicity effect. When the accretion process begins, the primary
star (WD) develops an optically thick wind which helps in stabilizing
the mass transfer process. When the metallicity is low ($[Fe/H] <
-1$), the stellar wind is too weak and the explosion cannot
occur. This model is appealing since it overcomes the difficulty of
achieving a stable accretion as in the classic SD scenario discussed
above. However, this model still needs to be tested in several
astrophysical contexts, as we will see in the next sections.

\end{itemize}

It is worth noting that in the last few years the DD scenario has lost
some credibility with respect to the SD scenario, mostly because of
the negative results of observational searches for very close binary
systems made of massive enough white dwarfs (Bragaglia et
al. 1990). Therefore, in this paper we will concentrate on the SD
model.

\subsection{Timescales}

In the framework of the SD scenario, the explosion timescales for
different progenitor models are given by the lifetime of the secondary
star. In particular:

\begin{itemize}

\item In the formulation of the type Ia rate by Greggio and Renzini
(1983b) (hereafter GR83), based on the Whelan and Iben (1973) model
(hereafter WI73) the explosion times correspond to the lifetimes of
stars in the mass range $0.8 - 8 M_{\odot}$. In fact, the maximum
initial mass which leads to the formation of a C-O WD is $\sim 8
M_{\odot}$, although stellar models with overshooting predict a lower
value (e.g. Marigo et al. 1996), which means that the first system,
made of two $8 M_{\odot}$ stars, explodes after $\sim 3 \cdot 10^7$
years from the beginning of star formation. The minimum mass of the
binary system is assumed to be 3 M$_{\odot}$, to ensure that the WD
and the companion are large enough to allow the C-O WD with the
minimum possible mass ($\sim 0.5M_{\odot}$, since stars with masses
below this limit can only become He-WDs) to reach the Chandrasekhar
mass after accretion. The smallest possible secondary mass is $0.8
M_{\odot}$ and therefore the maximum explosion time is the age of the
universe. This ensures that this model is able to predict a present
time SN Ia rate for those galaxies where star formation must have
stopped several Gyr ago, such as ellipticals.

\item Greggio (1996) (hereafter G96) revised the computation of the SN
Ia rate in the framework of the SD model and suggested a more detailed
criterium for the formation of a system which can eventually explode
as type Ia SN. In particular, the explosion will occur when:

\begin{equation}
M_{\rm WD}+\varepsilon M_{2, e}\geq M_{\rm Ch},
\end{equation}
\noindent
where $M_{2, e}$ is the envelope mass of the evolving secondary and
$\varepsilon$ is the accretion efficiency, namely the mass fraction of
the envelope which accretes over the WD. The maximum value for the
efficiency is $\varepsilon = 1$. The mass $M_{WD}$ is the mass of the
white dwarf: we have assumed three different $M_{in}-M_{WD}$
relationships, from Renzini and Voli (1981) (RV81), from Iben (1991)
(I91) and from Marigo et al. (1996; 1998) (MBC).

\item Kobayashi et al. (1998, 2000) (hereafter K98 and KTN) adopted Hachisu et
al. (1996; 1999) model and considered two possible progenitor systems:
either a WD plus a red giant (RG) star or a WD plus a main sequence
star (MS). In both cases the mass of the primary star is defined in
the range $3.0 - 8 M_{\odot}$, whereas the secondary masses are ($1.8
M_{\odot}\simlt M_{\rm MS}\simlt 2.6 M_{\odot}$) and ($0.9
M_{\odot}\simlt M_{\rm RG}\simlt 1.5 M_{\odot}$) for the MS and the RG
stars, respectively. In this case, the explosion times are given again
simply by the lifetimes of the secondary stars. If we compare this model
with the GR83 model, we see that the explosion timescales here are
much longer, since the most massive secondary is a $2.6 M_{\odot}$ star
with a lifetime of $\sim 3.3 \cdot 10^{8}$ years, instead of a $8 M_{\odot}$
with a lifetime of $\sim 3.0 \cdot 10^{7}$ years. 
In addition, the systems able to give rise
to SNe Ia may form only after the gas has reached a metallicity of
[Fe/H] $=-1.0$ dex (K98) or [Fe/H] $=-1.1$ dex (KTN),
and this introduces an additional delay which has
important consequences on galactic chemical evolution as we will see
in the next sections.

\end{itemize}

\subsection{Definition of $t_{Ia}$}

We adopt as definition of the {\it typical timescale for the type Ia SN
maximum enrichment} the time $t_{Ia}$ at which the SNIa rate reaches a
maximum. As we will see in the next sections this maximum depends upon
the assumed type Ia SN progenitor model, namely the lifetimes of the
progenitors and the initial mass function (IMF), and the star
formation rate (SFR). The combination of all these parameters creates
different shapes and different maxima in the type Ia SN rate versus
time relationship.

\section{THEORETICAL SN IA RATES}

In this section we describe the computation of the SN Ia rate
according to different histories of star formation: a) an
instantaneous burst (one stellar generation), b) constant star
formation rate, c) a more realistic star formation rate derived
in the framework of chemical evolution models.

\begin{itemize}

\item a) The calculation of the SNIa rate following a burst of star
formation of negligible duration, in the framework of the SD scenario
and the WI73 model, can be expressed as:

\begin{equation}
R_{\rm Ia}(t)=A \int_{M_{\rm B, inf}}^{M_{\rm B, sup}}\phi(M_{\rm B})
f\biggl({M_2(t)\over M_{\rm B}}\biggr){dM_{\rm B} \over M_B},
\end{equation}
where $M_B = M_1 + M_2$ is the total mass of the binary system,
$M_{\rm B, inf}$ and $M_{\rm B, sup}$ are the minimum and maximum
masses for the binary systems contributing at the time $t$. The
maximum value that $M_B$ can assume is called $M_{BM}$ and the minimum
$M_{Bm}$. These values (maximum and minimum mass of the binary system
able to produce a SNIa explosion) are model-dependent. In particular,
GR83 considered that only stars with $M \leq 8 M_{\odot}$ could
develop a degenerate C-O core, thus obtaining an upper limit $M_{BM} =
16 M_{\odot}$ for the mass of the binary system. The adopted lower
limit is ($M_{Bm} = 3 M_{\odot}$), as discussed previously. 
The extremes of the
integral (2), at a fixed time $t$, are:

\begin{equation}
M_{\rm B, inf} = max(2 M_2(t), M_{Bm})
\end{equation}

\begin{equation}
M_{\rm B, sup} = {1\over 2} M_{BM} + M_2(t),
\end{equation}
\noindent
where $M_{\rm B, sup} = M_{BM}$ when $M_2(t) = 8 M_{\odot}$.

We define $\mu=M_2/M_{\rm B}$ as the mass fraction of the secondary
and $f(\mu)$ is the distribution function of this ratio. Statistical
studies (e.g. Tutukov \& Yungelson 1980) indicates that mass ratios
close to one are preferred, so the formula:

\begin{equation}
f(\mu)=2^{1+\gamma}(1+\gamma)\mu^\gamma,
\end{equation}
\noindent
is commonly adopted (GR83), with $\gamma$=2 as a parameter. 
However, it is worth mentioning that in the literature there are different suggestions about the value of $\gamma$.
For example, Duquennoy and Mayor (1991) found a value of $\gamma$=-0.35, 
the value adopted in K98. Therefore, we will explore in this paper more values 
for the parameter $\gamma$ (see Table 1).

In order to
obtain the SNIa rate, we need also the function $M_2(t)$. To this
purpose we adopt the inverse of the formula:

\begin{equation}
\tau(M)=10^{[1.338-\sqrt{1.79-0.2232(7.764-\log(M))}]/0.1116}\,\rm yr,
\end{equation}
\noindent
(Padovani \& Matteucci 1993).

K98 adopted stellar lifetimes depending on the initial stellar metallicities from Kodama (1997): these timescales are systematically longer than ours especially in the domain of low and intermediate mass stars.

The function $\phi(M_{\rm B})$ is the IMF:

\begin{equation}
\phi(M_{\rm B}) \propto M_B^{-(1+x)}
\end{equation}
\noindent
where $x=1.35$ is the Salpeter (1955) index and the IMF is defined in
the mass interval 0.1-100 $M_\odot$. In the paper we have used both the
Salpeter and the Scalo(1986) IMF. Finally, the constant A, already
defined in GR83, is the fraction of binary systems, in the IMF, with the right
characteristics to become type Ia SNe and is fixed by reproducing the
present time observed type Ia SN rate.

For the G96 model the expression of the SN Ia rate is the same as
eq. (2) whereas the lower limit for the binary system $M_{\rm B, inf}$
at the time $t$ is:

\begin{equation}
M_{\rm B, inf}=max(2\,M_2(t), M_{\rm 1min}+M_2(t)),
\end{equation}
where $M_{\rm 1min}$ is the minimum mass the primary should have to
give rise to a WD whose mass satisfies eq. (1). We consider also a
Sub-Chandrasekhar model in which the explosion occurred when $M_{\rm
WD}\geq 0.6 \,M_\odot$ and $\varepsilon\cdot M_{\rm 2, e}\geq 0.15
\,M_\odot$. For this model we assume $\varepsilon=0.5$.

The calculation of the SNIa rate with KTN prescriptions is a little
more complicated, because we have different restrictions for the
primary and the secondary star: the secondary lies in a narrow range
of masses ($0.9 M_{\odot}\simlt M_2\simlt 1.5 M_{\odot}$ for the WD+RG
system and $1.8 M_{\odot}\simlt M_2\simlt 2.6 M_{\odot}$ for the WD+MS
system). Then, K98  and KTN consider only WDs with masses $0.7\,M_\odot \simlt
M_{\rm WD} \simlt 1.2\,M_\odot$, which corresponds to an interval for
the primary star $3\,M_\odot \simlt M_1 \simlt 8\,M_\odot$. In this
case the maximum and minimum values allowed for the progenitor binary
systems should be computed by taking into account the above
conditions, namely, $M_{BM}= 9.5 M_\odot$ and $M_{Bm}= 3.9 M_\odot$
for the case WD + RG and $M_{BM}= 10.6 M_\odot$ and $M_{Bm}= 4.8
M_\odot$ for the case WD + MS.
For the mass $M_{\rm B, inf}$ we have:

\begin{equation}
M_{\rm B, inf}=max(2\,M_2, M_{Bm}, M_{1min}+M_2)
\end{equation}
\noindent
and for the mass $M_{\rm B, sup}$:

\begin{equation}
M_{\rm B, sup}=min(M_{BM}, M_{1max}+M_2)
\end{equation}
\noindent
where $M_{1max}= 8 M_\odot$ and $M_{1min} = 3 M_\odot$.

In Figure 1 we show the SN Ia rate computed in the case of an
instantaneous burst for the GR83, G96 and KTN (without the metallicity
effect) prescriptions. As one can see, in the first two cases the rate
first increases, reaches a maximum at $t_{Ia} \sim 5 \cdot 10^7$ years
and then declines. This behaviour is due to the competition between
the increasing number of stars with decreasing mass (IMF) and the
increasing stellar lifetime with decreasing mass. The maximum in the
G96 cases is slightly anticipated relative to GR83, owing to the fact
that a maximum mass producing a C-O WD of 8.8 $M_\odot$ was
adopted. The other differences between the G96 ($\varepsilon=1$, 
$\gamma=0.5$) and
GR83 ($\gamma=0.5$) cases are due to condition 
(1) and the different $M_i - M_{WD}$
relations adopted. On the other hand, the G96 ($\varepsilon < 1$)
cases predict quite different SN Ia rates. In particular, as the
efficiency of mass transfer decreases, the number of systems able to
produce a type Ia SN is drastically reduced because of condition
(1). As a consequence of this, the case with $\varepsilon = 0.2$
predicts that type Ia SN rate following an instantaneous burst goes to
zero already after a time smaller than 1 Gyr.

The rate predicted by the KTN model shows instead a completely
different behaviour, with a discontinuity due to jump from 1.8
$M_\odot$ to 1.5 $M_\odot$ in the mass of the secondary passing from
the system WD + MS to WD + RG. In figure 1 the rates are normalized to
their own maximum value as in GR83, whereas in figure 2 we plot the
absolute SN rates normalized to reproduce a short starburst like those
occurring in blue compact galaxies. In particular, we assumed A=0.006,
a value which can reproduce the features of IZw18 (see Recchi et
al. 2001).

\item b) The single-burst approximation could be valid only for
starburst galaxies whereas for elliptical and spiral galaxies we must
consider a star formation extended in time. In the case in which the
SFR, $\Psi(t)$, is constant in time, we obtain:

\begin{equation}
R_{\rm Ia}(t)=A \Psi \int_{M_{\rm B, inf}}^{M_{\rm B, sup}}\phi(M_{\rm B})
\int_{\mu_{\rm min}}^{\mu_{\rm max}}f(\mu)d\mu\, dM_{\rm B},
\end{equation}
\noindent
Where $\mu$ is defined in the range $\mu_{\rm min}-0.5$, with:

\begin{equation}
\mu_{\rm min} = max({M_2 \over M_B}, {{M_B - 0.5 M_{BM}}\over M_B})
\end{equation}
\noindent
for GR83 and G96, whereas for KTN we define:

\begin{equation}
\mu_{\rm min} = max({M_2 \over M_B}, {({M_B - M_{1max})}\over M_B}, 
{M_{2min}\over M_B})
\end{equation}
\noindent
and:
\begin{equation}
\mu_{\rm max} = min({M_{2max} \over M_B}, {({M_B - M_{1min})}\over M_B})
\end{equation}
\noindent
The equation (11) is useful to calculate the SNIa rate in the GR83 and
G96 cases either for a burst with constant star formation rate or for
the solar neighbourhood where the SFR has not varied much in time,
whereas in all the other cases a constant SFR represent a poor
approximation. For a correct use of the assumption of K98 and KTN 
we must consider the
metal enrichment of the ISM (we cannot produce SNIa progenitors until
$[Fe/H] = -1.0$ (K98) or $-1.1$ (KTN). So, in order to calculate the 
SNIa rate, 
either with constant or variable SFR, we need a
chemical evolution model, as shown in c).

\item c) The case with a more realistic star formation rate first
increasing then reaching a maxiumum and decreasing, as predicted by
successful models for the chemical evolution of the solar neighbourhood
(e.g. Chiappini et al. 1997), can be written as:

\begin{equation}
R_{\rm Ia}(t)=A \int_{M_{\rm B, inf}}^{M_{\rm B, sup}}\phi(M_{\rm B})
\int_{\mu_{\rm min}}^{\mu_{\rm max}}f(\mu)
\psi(t-\tau_{M_2})d\mu\, dM_{\rm B},
\end{equation}
\noindent
as shown first by Matteucci \& Greggio (1986). The star formation rate
in this case has to be evaluated at the time $(t-\tau_{M_2})$, with
$\tau_{M_2}$ being the clock for the explosion. In figure 3 we show
the type Ia SN rates normalized to their maxima as obtained by
eq. (15) when the SFR of the Chiappini et al. (1997) model is
adopted. Such a SFR is proportional to a power $k = 1.5$ of the
surface gas density and to a power $h = 0.5$ of the total surface mass
density. Such a formulation of the SFR takes into account the feedback
mechanism between stars and gas regulating star formation and is
supported by observations (e.g. Dopita and Ryder 1994). We adopted two
chemical evolution models: i) the one of Matteucci and Fran\c cois
(1992) which is similar to that adopted by K98 and KTN. In this model the halo
and disk form out of a unique infall episode with a time scale for the
formation of the disk at the solar neighbourhood of $\tau = 4$
Gyr. ii) the model of Chiappini et al. (1997) which is known as {\it two
infall model} where the halo and thick disk are assumed to form on a
short timescale ($\sim$ 1 Gyr) out of a first episode of infall of
primordial gas, whereas the thin-disk is assumed to have formed on a
longer timescale ($\sim$ 8 Gyr at the solar circle). In this model,
the existence of a threshold in the gas density to regulate the star
formation (Kennicutt 1989; 1998) is also adopted. As a consequence of
this, the star formation rate goes to zero every time that the gas
density decreases below the threshold ($\sim 7
M_\odot\,pc^{-2}$). This model reproduces the majority of the features
of the solar vicinity and the whole disk and we consider it as the
{\it best model}. In all the models we adopted the SFR of Chiappini et
al. (1997) as described before, but only in one case we assume the
threshold in the gas density.

The model parameters are described in Table 1, where the second column
indicates if the chemical evolution model for the solar neighbourhood
has one or two infall episodes. The third column shows the efficiency
of star formation $\nu$ (i.e. the constant in the SFR, expressed in
Gyr$^{-1}$) and the fourth the adopted timescale (in Gyr) for the
formation of the disk in the solar region. The fifth column contains
the adopted values for the parameter $\gamma$ in the distribution
function of the mass ratios in binary systems.  The sixth column
indicates if a threshold in the gas density for the SFR has been
adopted or not. The seventh column shows the assumed prescriptions for
the type Ia SN progenitors in the various models.  The eighth, ninth
and tenth columns contain the values for A, IMF and $M_i- M_{WD}$
relation, respectively.  Finally, in the eleventh column is shown the
present time SN Ia rate as predicted by each model. The yields adopted
in all the models are the same: RV81 for low and intermediate mass
stars and Woosley and Weaver (1995) for massive stars.

As one can see, in figure 3 the SNIa rate as predicted with the KTN
prescriptions including the metallicity effect (model M1) shows a
maximum very late ($t_{Ia} \sim$ 8 Gyr), as it should be expected
given the nature of the assumed progenitors. It is worth noting that
the model parameters in M1 are the same adopted in the chemical
evolution model described in KTN. The model with one infall but the
prescriptions of GR83 (M3) predicts a maximum at around $t_{Ia}= 4 -5$
Gyr, whereas the two-infall model (M2) predictions are a bit more
complicated showing two maxima, the first one at $t_{Ia}=1.5$ Gyr, due
to the first infall episode and the associated star formation, 
and the second one between 4 and 5 Gyr due to the SN Ia systems born
at the beginning of the thin-disk formation. The one infall model with
G96 prescriptions (M4) presents a maximum at $t_{Ia} = 3$ Gyr. In
figure 3 is shown also the SN Ia rate as predicted by Matteucci (1994)
by adopting the GR83 prescription, for an elliptical galaxy with
$10^{11}\,M_\odot$ of luminous mass. In this case, the SFR is very
efficient ($\sim 10-15$ times more than in the solar neighbourhood)
and lasts only for $\sim 0.4$ Gyr. Therefore, the maximum of the SN Ia
rate occurs already at $t_{Ia} = 0.3$ Gyr. In figure 4 the absolute
type Ia SN rates, computed by means of the models of Table 1, are
presented. In order to compute the absolute rates we had to assume a
specific value for the constant A, which represents the fraction of
binary systems in the IMF developing a type Ia SN explosion. In
principle, the parameter A should be fixed by reproducing the observed
SN Ia rate but it should also assume reasonable values. In our best
model (M2), which adopts a Scalo (1986) IMF with two slopes
($x_1=1.35$ for $M<2M_\odot$ and $x_2 = 1.7$ for $M \geq 2 M_\odot$),
the adopted value is A = 0.05 and the predicted present time type Ia
SN rate is in good agreement with the observed one ($\sim 0.18$ SN 100
$yr^{-1}$, with $H_0 = 60\,Km\,sec^{-1}\,Mpc^{-1}$, Cappellaro et
al. 1999), as shown in Table 1. For M1, with the prescriptions
of KTN, we adopted, as in their paper, the Salpeter IMF with $A_{MS}=0.05$
and $A_{RG}=0.02$ (where $A_{MS}$ refers to the systems WD+MS and $A_{RG}$
refers to the systems WD+RG), in order to
be able to compare our chemical results with theirs. The predicted
present time SN Ia rate in this case is too low. We could obtain a
reasonable value for the SN Ia rate with the K98 and KTN prescriptions only by
assuming $A=0.3$, which is a very high and perhaps unrealistic value
for such a fraction. In fact, a fraction of 30 \% of binary systems
giving rise to type Ia SNe would imply a total fraction of interacting
binaries in the mass range producing the SNe larger than 50 \% (De
Donder and Vanbeveren 2001). The same problem arises with the G96
prescriptions, as we can see in figure 4 where the rates of type Ia SN,
as predicted by models M4 and M5, are shown. In fact, models M4 and M5
predict too low type Ia SN rates at the present time unless we assume
$A=0.3$. The differences between M4 and M5 are due only to the
different $M_i-M_{WD}$ relations adopted in the two cases.

\end{itemize}

\section{THE EFFECTS OF THE TYPE IA SN RATE ON CHEMICAL EVOLUTION}

In this section we discuss the impact of the different SN Ia rates we
have discussed up to now on the [O/Fe] versus [Fe/H] relationship,
which is often used in the literature to infer the timescale for SN Ia
enrichment. In figure 5 we show the data from Gratton et al. (2000)
compared with different model predictions (in all the models the
normalization is made with the observed solar values of Anders and
Grevesse, 1989) including those taken from the paper of K98. These
latter seem to be the best in reproducing the observed trend but this
result is not clear. In fact, the predictions of our M1 (see figure
5), which is the copy of KTN model, show a very different behaviour
which is, in our opinion, more understandable: in particular, it does
not show any change in slope in the [O/Fe] ratio over the whole range
of [Fe/H], since the first systems giving rise to type Ia SNe form
only when the gas reaches the metallicity [Fe/H]=-1.1, and therefore
they will contribute to the chemical enrichment much later than the
time corresponding to [Fe/H]=-1.1 dex and even [Fe/H]=-1.0 dex, as
indicated by the maximum of the SN Ia rate occurring at 8 Gyr.  We do
not understand how K98, who assumed [Fe/H]=-1.0 dex as the limiting
metallicity for the formation of the type Ia SN progenitors, can find
a change in slope right at this precise metallicity.  We conclude that
the type Ia SN rate suggested by K98 and KTN, besides giving too low
values for the present time rate, unless an enormous amount of binary
systems is assumed, cannot reproduce the [O/Fe] versus [Fe/H]
relationship. This relationship, in fact, is characterized
observationally by a marked change in slope occurring at around
[Fe/H]=-1.0 dex, which corresponds to the transition between halo and
disk metallicities. In our best model the change in slope occurs
roughly at this point and it corresponds to a galactic age of $\sim$ 1
Gyr. This timescale, in turn, corresponds to the first maximum reached
by the type Ia SN rate in the two-infall model and can be considered
as the typical timescale for the SN Ia enrichment to become
important. It is worth noting that in our best model the trend of the
[O/Fe] ratio for [Fe/H] $< $ 1.0 dex is not flat but it shows a slight
slope since the very first type Ia SNe start occurring already at 30
Myr from the beginning of star formation (GR83 model), as discussed
also by Chiappini et al. (1999).

In conclusion, the best prescriptions for progenitor models of type Ia
SNe in order to reproduce the observed type Ia SN rate and the
chemical evolution of the Galaxy is still the GR83 model.

\section{CONCLUSIONS}

In this paper we have explored various SN Ia progenitor models and the
resulting SN rates for different histories of star formation and we
have defined the typical timescale for the maximum chemical enrichment
from type Ia SNe in different galaxies. We have studied the effects of
the SD scenario, as originally proposed by WI73, with the recipes of
GR83 and G96 as well as the more recent SD model proposed by K98 and
KTN. Then we have calculated the chemical evolution of the solar
neighbourhood in the different cases by means of chemical evolution
models. Our results can be summarized as follows:

\begin{itemize}

\item The best prescriptions to obtain type Ia SN rates in agreement
with the observations seem to be those of GR83. If one adopts the more
realistic approach developed by G96 to calculate the rate of explosion
of systems made of a C-O WD plus a red giant star, then a large
fraction ($\sim$ 30 \%) of interacting binary systems giving rise to
type Ia SNe is required to reproduce the observed rate and the solar
Fe abundance. This corresponds to an overall interacting binary
frequency $> 50\%$ in the mass range of the primary star.
The same is true for the model K98 and KTN,
which presents also some other problems concerning galactic chemical
evolution.

\item In particular, the models of K98 and KTN do not produce satisfactory
results for the [O/Fe] vs. [Fe/H] relation in the solar neighbourhood
since it predicts that binary systems able to form type Ia SNe can
form very late, after the gas in the Galaxy has reached [Fe/H]=-1.1--  -1.0
dex. This fact clearly prevents the change in slope in the [O/Fe]
ratio right at [Fe/H] $\sim$ -1.0, as observed in the data. The
predicted metallicity effect seems also at variance
with recent observations of Damped Lyman-$\alpha$ systems
(e.g. Pettini et al. 1999) indicating almost solar [$\alpha$/Fe]
ratios in objects with metallicities lower than [Fe/H]=-1.0 dex, thus
suggesting that type Ia SNe have already polluted the interstellar
medium at low metallicities.
However, KTN suggested 
a possible explanation for this effect. In particular, they 
attributed the almost solar ratios at low metallicities as due to 
inhomogeneous chemical evolution in those early phases.
In our opinion this solution looks rather ``ad hoc''
and is not justifiable for metallicities [Fe/H] $> -3.0$ dex, since no 
large spread is observed for these metallicities in the 
abundance ratios of halo stars (Argast et al. 2000)

\item The typical timescale for the maximum chemical enrichment by SN
Ia, $t_{Ia}$, can be defined as the time at which the SN Ia rate
reaches a maximum. This timescale depends upon the adopted SN
progenitor model, the stellar lifetimes, the IMF and the SFR. Since
galaxies of different morphological type are likely to have different
histories of star formation, the typical type Ia SN timescale are
necessarily different from galaxy to galaxy.

\item We have studied different cases in the framework of GR83 model
for SN Ia progenitors: in an instantaneous starburst, which can
approximate the situation of blue compact galaxies, $(t_{Ia})_{BC}
\sim 40 - 50$ Myr, in a spiral galaxy like the Milky Way, where the
SFR has not varied much in time but it has been more ore less
continuous, $(t_{Ia})_{Sp} \sim 4 - 5$ Gyr. A minor maximum appears at
$\sim$ 1 Gyr in the two-infall model, where the halo is explicitely
taken into account and which better fits the properties of the solar
neighbourhood, corresponding to the bulk of type Ia SNe originating
from the star formation in the halo. In an elliptical galaxy, where
the SFR was quite high but lasted for a relatively short time ($\sim
0.2 - 0.4$ Gyr), $(t_{Ia})_{E} \sim 0.3$ Gyr.

\item The knee observed in the [$\alpha$/Fe] vs. [Fe/H] relations in
the solar neighbourhood stars corresponds to a timescale $t_{knee} =
1$ Gyr. This is the time at which the Fe production from type Ia SNe
starts to become important, due to the systems formed during the halo
phase, and is not a universal value but peculiar to the solar region.

\item The type Ia SN rates per unit luminosity
were plotted as function of redshift (see
figure 6), in a cosmology where we assume that all the galaxies formed
at $z_f=5$, $\Omega_{\rm M}=0.3$ and $\Omega_{\Lambda}=0.7$. The
Hubble constant is assumed to be H$_0 = 60$ Km s$^{-1}$ Mpc$^{-1}$. 
The photometric model we adopted is that of Jimenez et al. (1998)
As
one can see in figure 6, where the rates for a typical elliptical and
a typical spiral galaxy, expressed in SNu, are shown, a maximum at a redshift
$z$ between 3 and 4  and extending until $z \sim 1.6$
should be expected for the elliptical galaxies, whereas we
predict two peaks for spiral galaxies (two-infall model) the first at
$z \sim 3$ during the halo phase and the second at $z \sim 1$, 
during the disk
phase. 
In the same figure are
indicated the present time values for the type Ia rate 
in ellipticals and spirals as well as the type Ia rate derived at
redshift $z \sim 0.4$ by Pain et al. (1996). The observed values have
been scaled to our adopted $H_0$. The agreement between the observed
and predicted rates seems quite good (especially for the present time
rates), although it would be more appropriate to compute the cosmic SN
rate (i.e. weighted over the different morphological types) in order
to compare it with the high redshift observations.
\end{itemize}

\acknowledgements We are indebted to Laura Greggio for many
discussions and for providing her code to compute the SN rates in the
case of a starburst. We also thank Renato Manara and Cristina
Chiappini for their contribution to this work and S. Borgani for
reading the manuscript. 
We thanks the suggestions of an anonymous referee which 
much improved the final version of this paper.
This work has been supported by research funds
from M.U.R.S.T. (Cofin 1998).

\clearpage

\begin{landscape}
\begin{table*}
\begin{flushleft}
\caption[]{Model Parameters and Predicted SN Ia rates}
\begin{tabular}{||c||c|c|c|c|c|c|c|c|c|c||}
\noalign{\smallskip}
\hline
\noalign{\smallskip}
Model & Infall & $\nu$ & $\tau$ & $\gamma$ & Threshold & SNeIa & A & IMF & 
$M_i - M_{WD}$ & Rates (100$yr^{-1}$)\\
\noalign{\smallskip}
\hline
M1 & 1 & 0.37 & 5 & -0.35 & No & KTN & 0.05 WD+MS & Salpeter$^*$ 
& RV81 & 0.025\\
 & & & & & & & 0.02 WD+RG & & &\\
M2 & 2 & 2; 1 & 8 & 2 & Yes & C97 & 0.05 & Scalo & RV81 & 0.172\\
M3 & 1 & 0.5 & 4 & 2 & No & C97 & 0.05 & Scalo & RV81 & 0.193\\
M4 & 1 & 0.5 & 4 & 0.5 & No & G96 & 0.05 & Scalo & I91 & 0.039\\
M5 & 1 & 0.5 & 4 & 0.5 & No & G96 & 0.05 & Scalo & MBC & 0.023\\
\hline
\noalign{\smallskip}
\end{tabular}

\medskip

* defined in the range 0.05-50 $M_\odot$
\end{flushleft}
\end{table*}
\end{landscape}

\newpage
\begin{figure}
\figurenum{1}
\epsscale{0.9}
\plotone{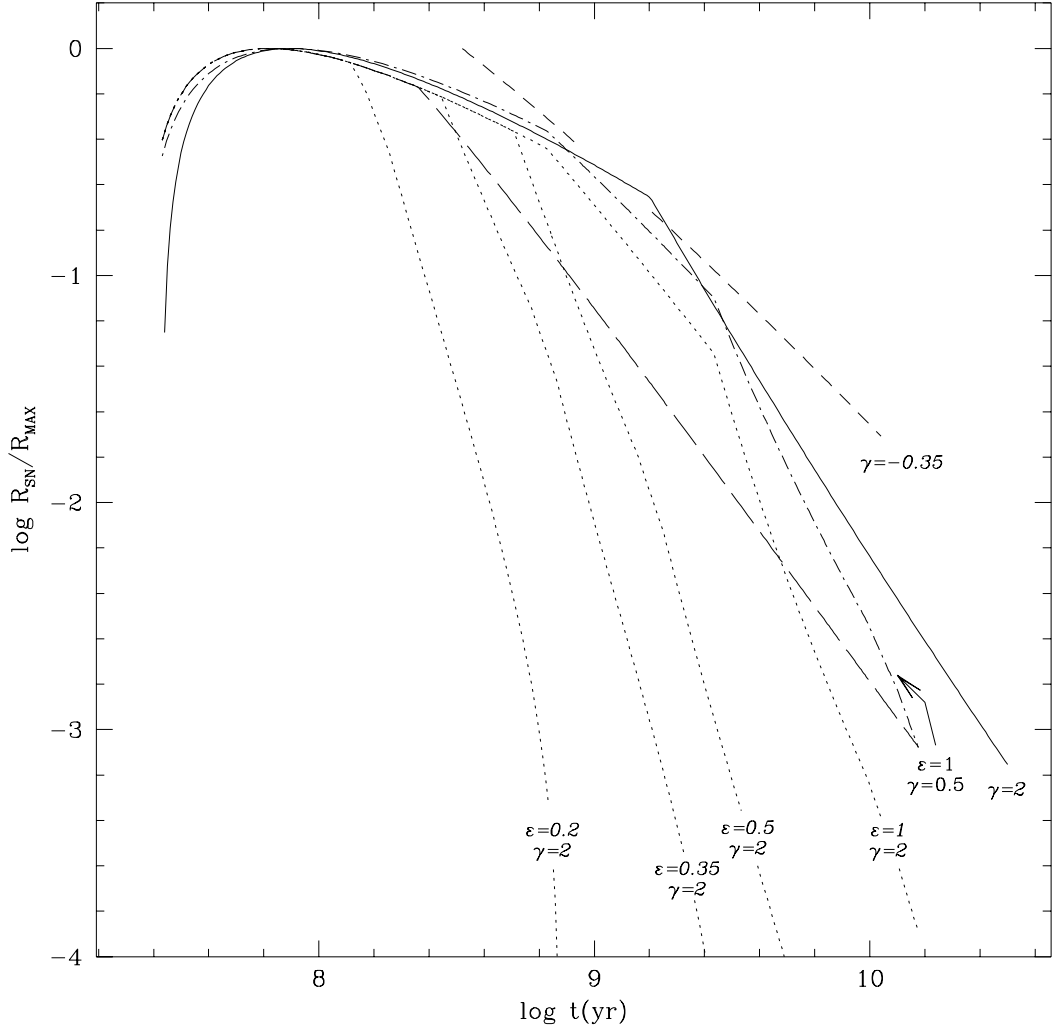}
\caption[f1.eps]{\label{f:1} SNIa rates vs. time for an
instantaneous burst of star formation. These rates are normalized to
their respective maximum values. The solid line is the GR83 model. The
dotted lines are the G96 models with different values of $\varepsilon$
(0.2, 0.35, 0.5 and 1 respectively), the long-dashed line is the
Sub-Ch G96 model ($\varepsilon=0.5$) and the short-dashed lines are
the K98 model without the metallicity effect. In all the models we
have assumed a Salpeter IMF and $\gamma=2.0$, except in the K98 model
where we assumed $\gamma=-0.35$ in order to reproduce exactly their
model.  A model with G96 prescriptions and $\gamma=0.5$ is also shown
(dot-dashed line).  The $M_i - M_{WD}$ relationship for the solid line
is from RV81 whereas for the G96 model is from I91.  }
\end{figure}

\begin{figure}
\figurenum{2}
\epsscale{0.9}
\plotone{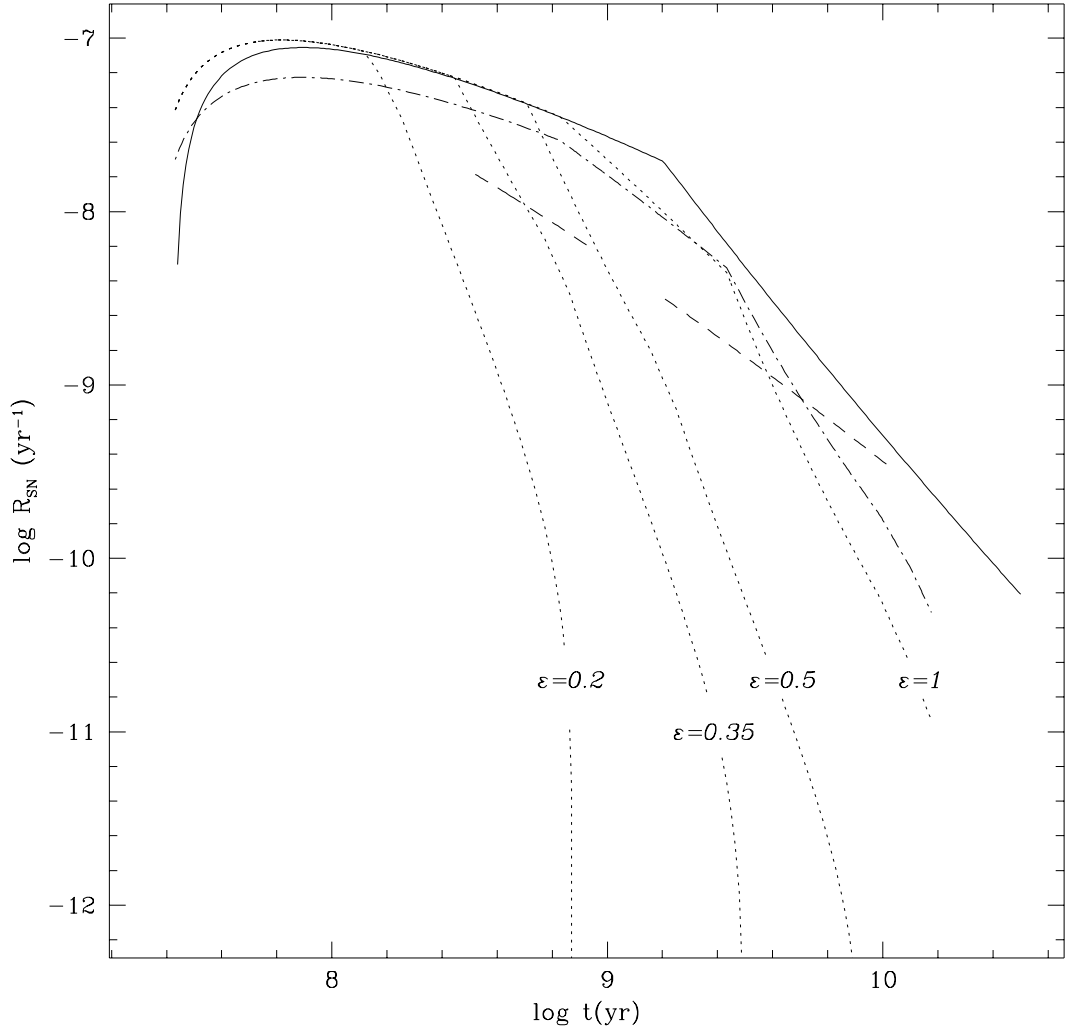}
\caption[f2.eps]{\label{f:2} Same as Fig. 1, but with the absolute
values of SNIa rates, in units of SNe $yr^{-1}$.  The assumed value of
A=0.006 is from Recchi et al. (2001) and is chosen to fit the
properties of the galaxy IZw18.  }
\end{figure}

\begin{figure}
\figurenum{3}
\epsscale{0.9}
\plotone{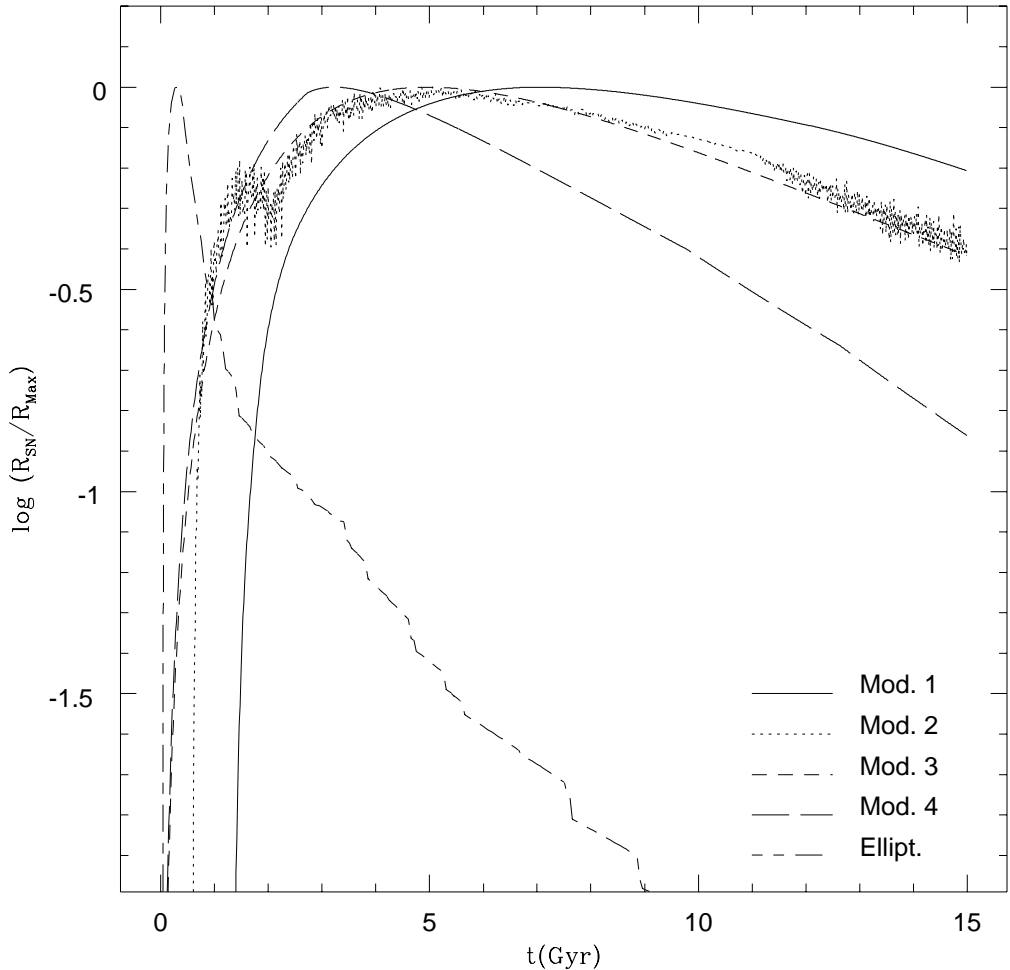}
\figcaption[f3.eps]{SNIa rates normalized to their maxima and
computed by means of a star formation rate depending on the gas density in
the framework of the chemical evolution models shown in Table 1. The
continuous line corresponds to M1, the short-dashed line to M3 and the line
with the small oscillations to M2, where a threshold in the gas density
for the star formation is assumed. The long-dashed line corresponds
to model M4. The short-dashed-long-dashed line is the rate predicted for an
elliptical galaxy by Matteucci (1994).
\label{f:3}}
\end{figure}

\begin{figure}
\figurenum{4} 
\epsscale{0.9}
\plotone{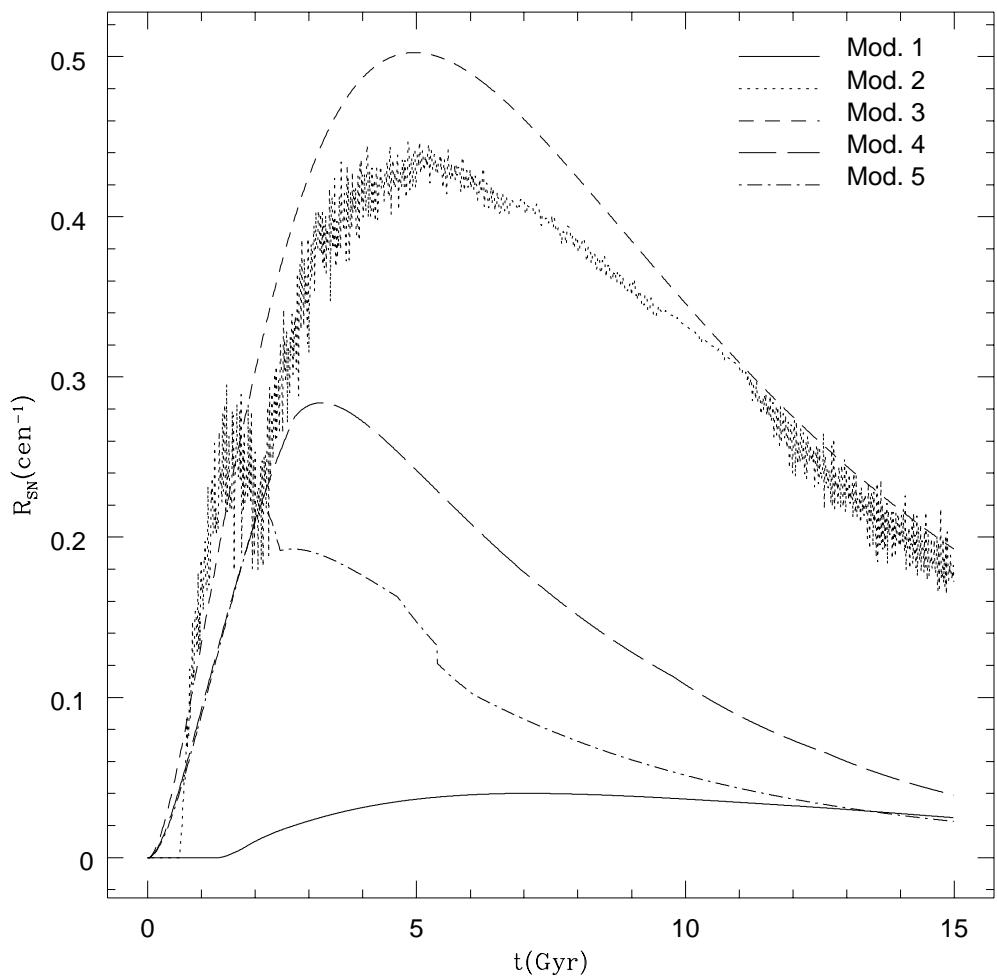} 
\figcaption[f4.eps]{The absolute SN
rates in units of SNe $yr^{-1}$ computed for the models described in
Table 1, where all the assumed model parameters are shown. The symbols
are the same as in fig.3. We report here also the prediction of Model
5 (dot-dashed line).  The G96 models (Mod. 4 and Mod. 5) are computed
by adopting the $\varepsilon =1$.
\label{f:4}}
\end{figure}

\begin{figure}
\figurenum{5}
\epsscale{0.9}
\plotone{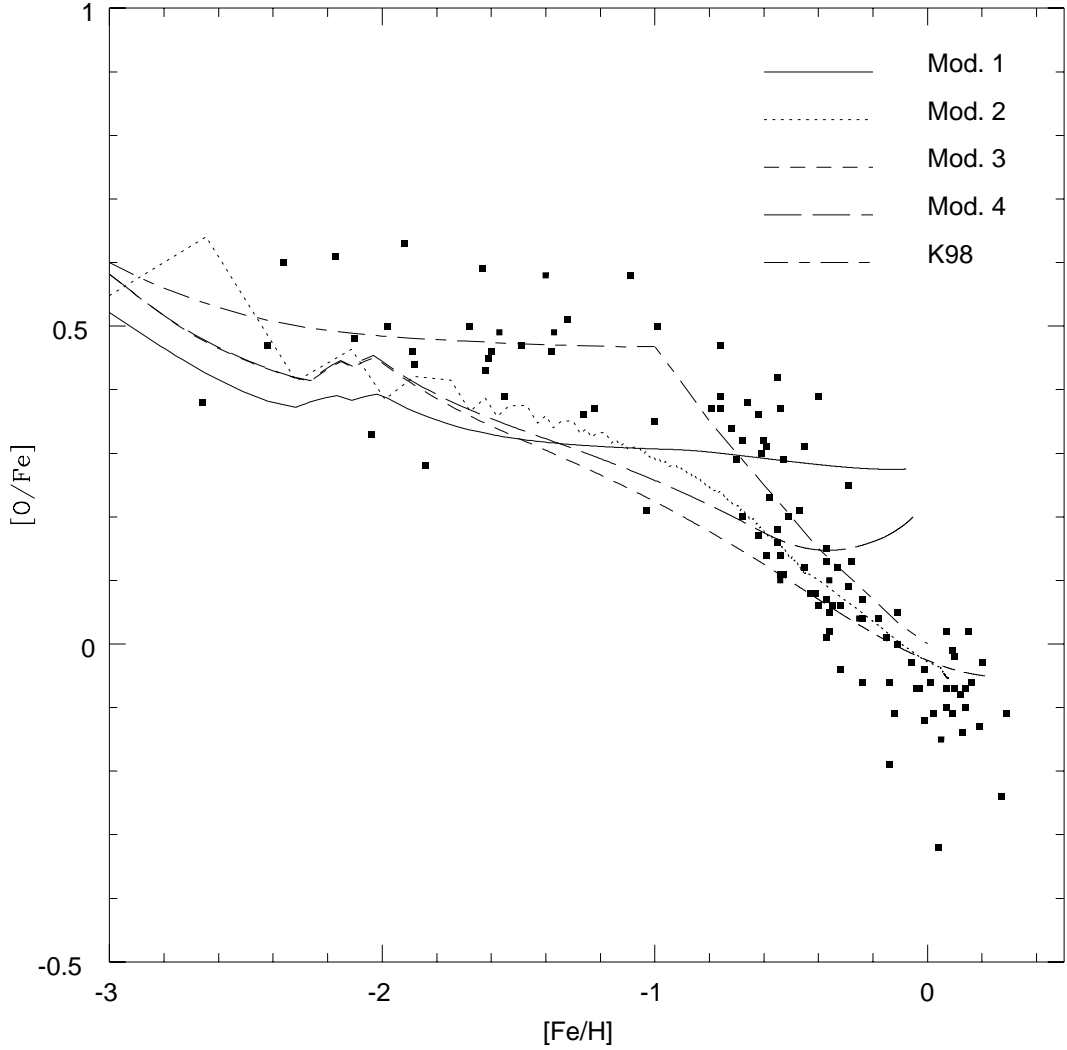}
\figcaption[f5.eps]{The observed and predicted [O/Fe]
vs. [Fe/H] relationships. The predictions are from the models of Table
1. The abundances are all normalised to the observed solar abundances
(Anders and Grevesse, 1989).
The dashed-dotted line is the model computed by K98, their
figure 3, that we show for comparison. 
The data points (filled squares) are from Gratton et al. (2000).
\label{f:5}}
\end{figure}

\begin{figure}
\figurenum{6}
\epsscale{0.9}
\plotone{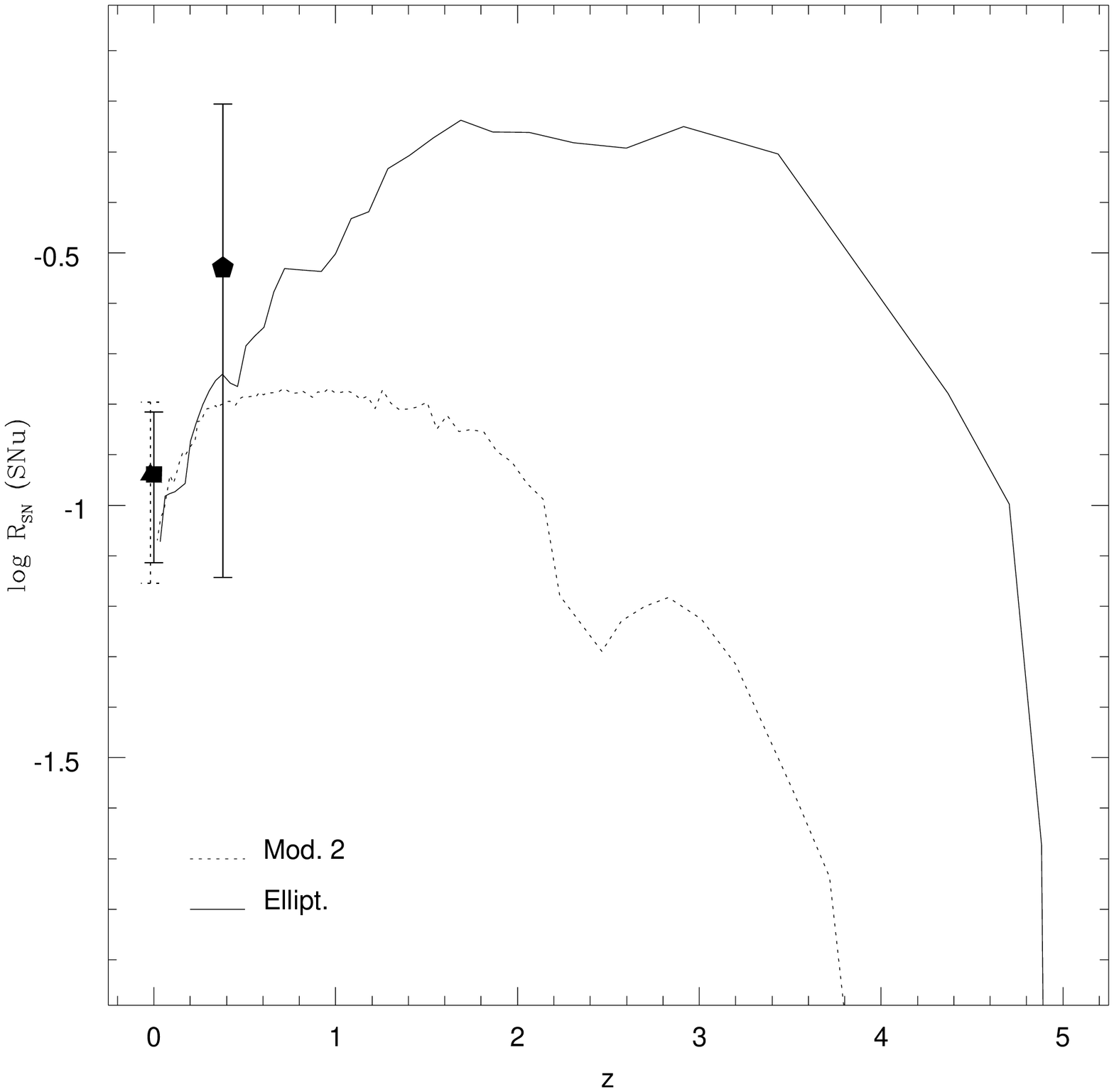}
\figcaption[f6.eps]{Predicted type Ia SN rates expressed in
SNu ($SNe 100 yr^{-1} 10^{-10} L_{B_\odot}$) in different galaxies as 
functions of redshift. The continuous line represents an elliptical
galaxy (the same as figure 3) and the dotted line indicates a spiral
(model M2). The observed local type Ia SN rates in ellipticals (the
square) and spirals (the triangle)  
together with the rate measured by
Pain et al. (1996) at $z \sim 0.4$ (the pentagon) are shown.
\label{f:6}}
\end{figure}


\begin{thebibliography}{99}


\bibitem{ag} Anders, E., Grevesse, N. 1989 Geochim. Cosmochim. Acta 53, 197
\bibitem{ar} Argast, D., Samland, M., Gerhard, O.E., Thielemann, F.-K. 2000,
A \& A 356, 873
\bibitem{bg} Bragaglia, A., Greggio, L., Renzini, A., d'Odorico, S. 
1990, Ap.J. 365, L13
\bibitem{ce} Cappellaro, E., Evans, R., Turatto, M. 1999, A \& A 459, 466
\bibitem{c1} Chiappini, C., Matteucci, F., Gratton, R.G. 1997, 
Ap.J. 477, 765
\bibitem{c2} Chiappini, C., Matteucci, F., Beers, T.C., Nomoto, K. 
1999, Ap.J. 515, 226
\bibitem{dv} De Donder, E., Vanbeveren, D. 2001, A \& A in press
\bibitem{dr} Dopita, M.A., Ryder, S.D. 1994, Ap.J. 430, 163
\bibitem{dm} Duquennoy, A., Mayor, M. 1991, A \&A 248, 485
\bibitem{gc} Gratton, R.G., Carretta, E., Matteucci, F., Sneden, C. 
2000, A \& A 358, 671
\bibitem{gr1} Greggio, L., Renzini, A. 1983a, Mem. SAIt 54, 311
\bibitem{gr2} Greggio, L., Renzini, A. 1983b, A \& A 118, 217 (GR83)
\bibitem{g} Greggio, L. 1996, in ``The Interplay between Massive Star 
Formation, the ISM and Galaxy Evolution'' ed. D. Kunth et al. Edition 
Frontieres, p.98 (G96)
\bibitem{kh} Kenyon, S.J., Hartmann, L., Gomez, M., Carr, J.S.,
Tokunaga, A. 1993, A.J. 105, 1505
\bibitem{k1} Kennicutt, R.C., 1989, Ap.J. 344, 686
\bibitem{k2} Kennicutt, R.C., 1998, Ap.J. 498, 541
\bibitem{k98} Kobayashi, C., Tsujimoto, T., Nomoto, K., Hachisu, I.,
Kato, M. 1998, Ap.J. 503, L155 (K98)
\bibitem{kt} Kobayashi, C., Tsujimoto, T., Nomoto, K. 2000, Ap.J. 539, 26
(KTN)
\bibitem{k} Kodama, T. 1997, PhD Thesis, University of Tokyo
\bibitem{hk1} Hachisu, I., Kato, M., Nomoto, K. 1996, Ap.J. 470, L97
\bibitem{hk2} Hachisu, I., Kato, M., Nomoto, K. 1999, Ap.J. 522, 487
\bibitem{i} Iben, I. Jr, 1991, Ap.J. Suppl. 76, 55 (I91)
\bibitem{it1} Iben, I. Jr, Tutukov, A.V. 1984, Ap.J. Suppl. 54, 335
\bibitem{it2} Iben, I. Jr, Tutukov, A.V. 1985, Ap.J. Suppl. 58, 661
\bibitem{jpm} Jimenez, R., Padoan, P., Matteucci, F., Heavens, A.F. 1998, 
MNRAS
299,123
\bibitem{lt} Limongi, M., Tornamb\'e, A., 1991, Ap.J. 371, 317
\bibitem{mb1} Marigo, P., Bressan, A., Chiosi, C. 1996, A \& A 313, 545
\bibitem{mb2} Marigo, P., Bressan, A., Chiosi, C. 1998, A \& A 331, 564 (MBC)
\bibitem{m1} Matteucci, F., 1991 in ``SN 1987A and Other Supernovae'', ed. 
I.J. Danziger and K. Kj\"ar, ESO Publ., p.703
\bibitem{m2} Matteucci, F., 1994, A \& A 288, 57
\bibitem{mf} Matteucci, F., Fran\c cois, P. 1992, A \& A 262, L1
\bibitem{mg} Matteucci, F., Greggio, L. 1986, A \& A 154, 279
\bibitem{mt1} Matteucci, F., Tornamb\'e, A. 1987, A \& A 185, 51
\bibitem{mt2} Matteucci, F., Tornamb\'e, A. 1988, Comments on 
Astrophys. 12, 245
\bibitem{mr} Munari, U., Renzini, A. 1992, Ap.J. 397, L87
\bibitem{nt} Nomoto, K., Thielemann, F.K., Yokoi, K. 1984, Ap.J. 286, 644
\bibitem{p} Pain, R. et al. 1996, Ap.J. 473, 356
\bibitem{pe} Pettini, M., Ellison, S.L., Steidel, C.C., Bowen, D.V. 
1999, Ap.J. 510, 576
\bibitem{rm} Recchi, S., Matteucci, F., D'Ercole, A. 2001, MNRAS 322, 800
\bibitem{rv} Renzini, A., Voli, M. 1981, A \& A 94, 175 (RV81)
\bibitem{s} Salpeter, E.E., 1955, Ap.J. 121, 161
\bibitem{sc} Scalo, J.M. 1986, Fund. Cosmic Phys. 11, 1
\bibitem[Tinsley (1980)]{t80} Tinsley, B.M. 1980, Fund. Cosmic Phys. 5, 287
\bibitem{tm} Tornamb\'e, A., Matteucci, F. 1987, Ap.J. 318, L25
\bibitem{ty} Tutukov, A.V., Yungelson, L.B. 1980, in ``Close Binary 
Stars'', ed. M. Plavec et al. Reidel (Dordrecht), p.15
\bibitem{yt} Yoshii, Y., Tsujimoto, T., Nomoto, K. 1996, Ap.J. 462, 266
\bibitem{wi} Whelan, J., Iben, I. Jr. 1973, Ap.J. 186, 1007 (WI73)
\bibitem{ww} Woosley, S.E., Weaver, T.A. 1995, Ap.J. Suppl. 101, 181
\end{thebibliography}
\end{document}